\documentclass[letterpaper, 10 pt, conference]{ieeeconf}  
\IEEEoverridecommandlockouts                         

\usepackage{graphics} 
\usepackage{mathptmx}
\usepackage{amsmath} 
\usepackage{amssymb} 
\usepackage{graphicx}
\usepackage{bbold}

\usepackage{tabularx}

\usepackage{subcaption}

\usepackage{array}
\newcolumntype{P}[1]{>{\centering\arraybackslash}p{#1}}
\newcolumntype{M}[1]{>{\centering\arraybackslash}m{#1}}

\newtheorem{fact}{Fact}
\newtheorem{definition}{Definition}

\begin{document}

\title{\LARGE \bf
A Study of Three Influencer Archetypes for the Control of\\
Opinion Spread in Time-Varying Social Networks
}

\author{Michael DeBuse and Sean Warnick$^{\dagger}$
\thanks{$^{\dagger}$Michael DeBuse and Sean Warnick are with the Information and Decision Algorithms Laboratories (IDeA Labs), Department of Computer Science, Brigham Young University, Provo, UT 84602, USA {\tt\small mdebuse3@gmail.com}, and
{\tt\small sean@cs.byu.edu}}}

\maketitle
\thispagestyle{empty}
\pagestyle{empty}

\begin{abstract}
In this work we consider the impact of information spread in time-varying social networks, where agents request to follow other agents with aligned opinions while dropping ties to neighbors whose posts are too dissimilar to their own views.  
Opinion control and rhetorical influence has a very long history, employing various methods including education, persuasion, propaganda, marketing, and manipulation through mis-, dis-, and mal-information. The {\em automation} of opinion controllers, however, has only recently become easily deployable at a wide scale, with the advent of large language models (LLMs) and generative AI that can translate the quantified commands from opinion controllers into actual content with the appropriate nuance.   Automated agents in social networks can be deployed for various purposes, such as breaking up echo chambers, bridging valuable new connections between agents, or shaping the opinions of a target population---and all of these raise important ethical concerns that deserve serious attention and thoughtful discussion and debate.  This paper attempts to contribute to this discussion by considering three archetypal influencing styles observed by human drivers in these settings, comparing and contrasting the impact of these different control methods on the opinions of agents in the network. We will demonstrate the efficacy of current generative AI for generating nuanced content consistent with the command signal from automatic opinion controllers like these, and 
we will report on frameworks for approaching the relevant ethical considerations.

\end{abstract}

\section{INTRODUCTION}
Social media is a rich platform to share ideas and opinions, debate,  argue, and influence others. The use of automated agents, or {\it bots}, in such environments, however, raises important ethical issues related to the morality of persuasion.

The study of social interactions and rhetoric has a long history, dating at least as far as the ancient Greeks \cite{aristotleRhetoric}.
In the twentieth century, quantified methods for describing social relations were developed with the advent of {\it sociometry} \cite{Moreno1934Sociometry}, leading to the concept of a {\it social network}, {\it Social Network Analysis} (SNA) \cite{scott1988trend, wasserman1994SNA}, and {\it network science} \cite{strogatz2001exploring}, while an emphasis on {\it dynamics} and {\it control} for these systems began with Weiner's {\it Cybernetics} and its specialization as {\it sociocybernetics} \cite{wiener2019cybernetics, buckley1967sociology}. 
Nevertheless, according to \cite{proskurnikov2017tutorial, proskurnikov2018tutorial}, ``The realm of social systems has remained almost untouched by modern control theory in spite of the tremendous progress in control of complex large-scale systems."

Since 2017, when this observation was made, significant advancements have been made.  Leveraging diffusion and epidemic models \cite{jackson2005diffusion, lagnier2013predicting, jiang2014diffusion, wang2014studying, d2015interests}, controls researchers have modeled the {\em spread} of opinions \cite{pare2020network, pare2022network}.  Meanwhile, other researchers have focused on foundational models of {\em opinion formation} over static networks \cite{proskurnikov2017tutorial, taylor1968towards, degroot1974reaching, friedkin1990social}, eventually leading to models of opinion dynamics over state-dependent graphs or other time-varying network models \cite{proskurnikov2018tutorial, deffuant2000mixing, hegselmann2002opinion}.  Convergence and stability proofs have borrowed heavily from the consensus, flocking, and multi-agent systems literature \cite{olfati2004consensus, olfati2006flocking, cucker2007emergent, tanner2007flocking, olfati2007consensus}, and changes in these properties in the presence of certain types of agents, such as stubborn agents \cite{ yildiz2013binary, ijcai2017p124, hu2017competing, hunter2018optimizing, brooks2020model}, has lead to explicit work on network control \cite{wendt2019control}.  

\begin{figure}[t!]
      \centering
            \includegraphics[width=3in]{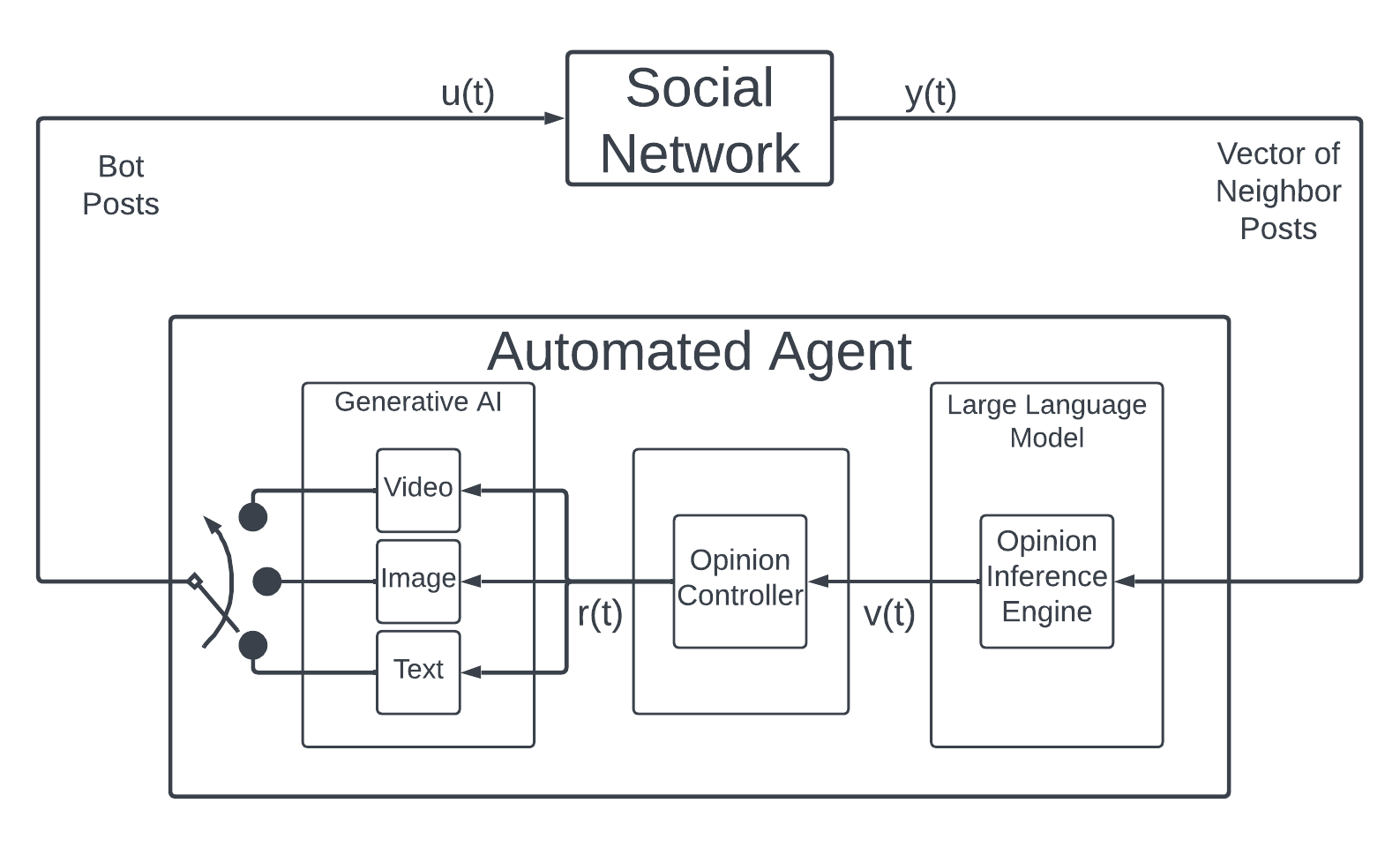}
      \caption{The feedback control of Social Networks using Automated Agents driven by Opinion Controllers coupled with Large Language Models and other Generative AI.}
      \label{fig:ControlDiagram}
\end{figure}

\begin{figure}[t]
      \centering
      \framebox{\parbox{3.3in}{
            \includegraphics[width=3.3in]{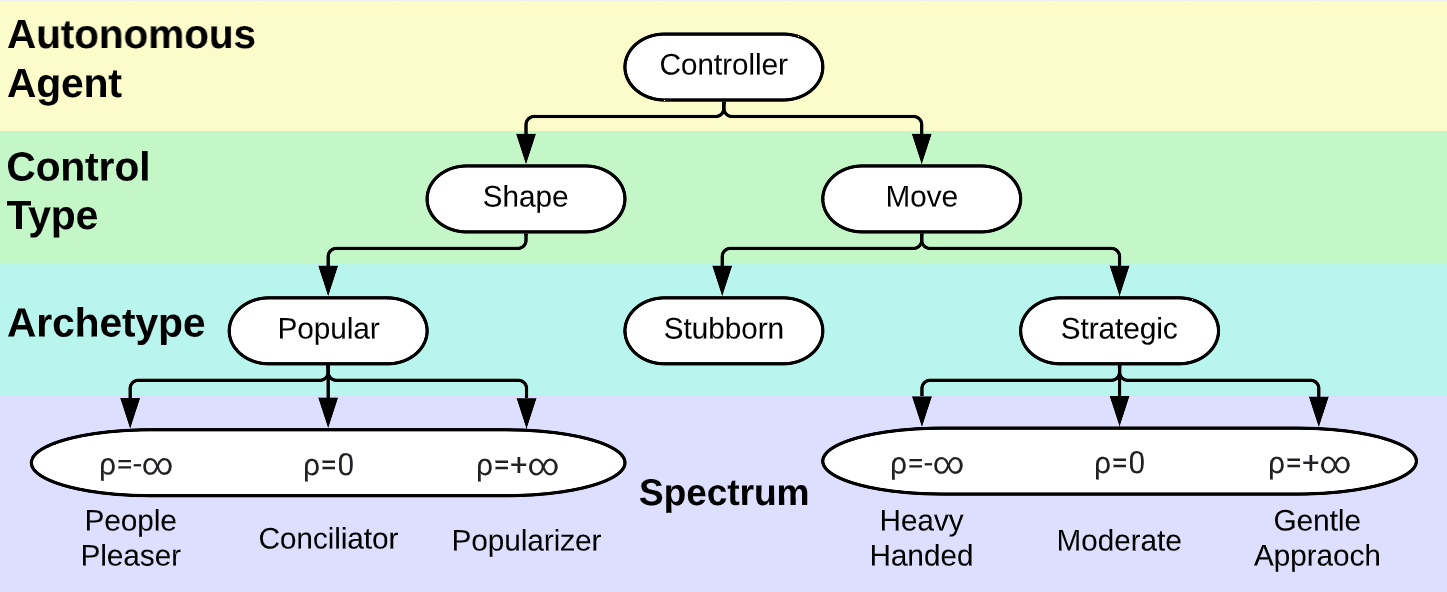}
        }}
      \caption{Tree showing the relationships between the three controller archetypes: Popular, Stubborn, and Strategic. The Hadamard power value, $\rho$, creates a spectrum of behaviors for the popular and strategic agent archetypes.}
      \label{fig:3archetypes}
\end{figure}

This paper builds on the results of these works, as well as many cited by these and other papers.  The contributions of this work include (see \mbox{Figure \ref{fig:ControlDiagram}}):
\begin{enumerate}
    \item Section \ref{sec:model}: Introduction of a novel nonlinear, stochastic Social Network model for the co-evolution of opinion and graph-topology dynamics,
    \item Section \ref{sec:controllers}: Novel models of archtypical classes of influencer dynamics observed in social networks as Opinion Controllers (see \mbox{Figure \ref{fig:3archetypes}}),
    \item Section \ref{sec:results}: Simulation studies of the archtypical controllers from Section \ref{sec:controllers} (see \mbox{Figures \ref{fig:initial_network}, and \ref{fig:combined_figures}}),
    \item Section \ref{sec:LLM}: Demonstration of the use of generative AI technologies for both a) Opinion Inference, quantifying the opinion reflected by content with respect to a list of topics and including results from associated validation studies (see Figure \ref{fig:post_rankings}), and b) Content Generation, converting the numerical control signal from
    the opinion controller into corresponding content that the
    Automated Agent can post, using video, images, and/or text---all making the realization and practical implementation of opinion controllers as Automated Agents in social networks very easy (see \mbox{Table \ref{tab:chatGPT} and Figure \ref{fig:climate_both}}), 
    \item Section \ref{sec:ethics}: A discussion of the ethical issues surrounding social control theory, including ethical frameworks from related areas for consideration. 
\end{enumerate}    

\section{Network Model}\label{sec:model}

The social network is represented as graph, $G=(V,E)$ with a set of $n$ vertices, $V$, that represent agents (users on the network) and a set, $E$, of ordered pairs of vertices representing directed edges, $E=\{(v_i,v_j)|$ \mbox{$v_i,v_j\in V$}\}.
This structure can be effectively represented by an adjacency matrix $A$, where entry $a_{ij}=1$ if $(v_j,v_i)\in E$ and $a_{ij}=0$ otherwise. 

We will consider time-varying networks, with a fixed number of vertices, representing agents or users of the social media platform, but where edges may appear or disappear as represented by $A_{[k]}$, $k = 0, 1,2,...$.  Moreover, we will assume $A_{[k]}$ is symmetric for all $k$, suggesting that if one agent ``follows" another on the social media platform, then the second agent will reciprocate by also following the first; edges thus denote a ``connection" between vertices.  Because of this symmetry, $G$ will always be undirected, as in \mbox{Figure \ref{fig:initial_network}}.

 Associated with each agent (i.e. vertex) is a vector of opinions on $m$ distinct topics, represented by an opinion matrix $X_{[k]}\in\mathbb R^{n\times m}$ with entries $0\leq x_{ij[k]}\leq 1$ indicating the degree of support agent $i$ feels towards topic $j$ at time $k$.  The $i^{th}$ row of $X_{[k]}$ is indicated by $x_{i[k]}^\top$, where $x_{i[k]}\in\mathbb R^m$, and the $j^{th}$ column of $X_{[k]}$ is indicated by $x_{j[k]} \in\mathbb R^n$.

To model opinion dynamics on the network, we will use a variation of the state-dependent French-DeGroot model \cite{proskurnikov2017tutorial,proskurnikov2018tutorial} given by:
\begin{equation}\label{eq:newX}
    X_{[k+1]} = W(X_{[k]},A_{[k]})X_{[k]}
\end{equation}
where $W(X_{[k]},A_{[k]})\in\mathbb R^{n\times m}$ is a row-stochastic weighting matrix that depends on both the {\it evolving} network opinions, $X_{[k]}$, and the {\it evolving} topology of the social network, $A_{[k]}$.

The novelty of this model comes from the way $W$ is calculated and the way the graph topology, $A_{[k]}$ {\it co-evolves} with agent opinions, $X_{[k]}$.  We will first describe the calculation of $W$, and then give the update equation for $A_{[k+1]}$.  

\begin{definition}
    Let $\epsilon > 0$ be a vanishingly small number; diag($v$) be a square, diagonal matrix with the entries of the vector $v$ on the diagonal; and $\mathbb 1$ be the (appropriately sized) vector of ones.  Then the {\it row-normalization operator} of a matrix $M$, ${\cal R}(M)$, is given by:
    \[
    {\cal R}(M) := {\rm diag}(M{\mathbb 1}+\epsilon{\mathbb 1}))^{-1}M
    \]
\end{definition}

\begin{definition}
    The {\it row-wise difference matrix} of a matrix $M$ is a square, symmetric, non-negative, hollow, row-stochastic matrix $D_N$ characterized by:
    \[
    D_N(M) := {\cal R}(D(M)),
    \]
    where $D(M)$ is a matrix with entries given by:
    \[
    d_{ij}(M) =  \|m_i^\top - m_j^\top\|_1. 
    \]
\end{definition}

\begin{definition}
    With $\mathbb 1$ and $I$ being the appropriately sized matrix of ones and identity matrix, respectively, the {\it row-wise similarity matrix} of a matrix $M$ is a square, symmetric, non-negative, hollow, row-stochastic matrix $S_N$ characterized by:
    \[
    S_N(M):= {\cal R}({\mathbb 1}-(I+D_N(M)))
    \]
\end{definition}

With these definitions, and noting that here $\mathbb 1$ and $I$ are the appropriately sized vector of ones and identity matrix, respectively, and $M_1\circ M_2$ is the element-wise, or Hadamard, multiplication of two appropriately sized matrices, $W$ can now be characterized:
\begin{equation}\label{eq:weights}
W(X_{[k]},A_{[k]}):=S_N(X_{[k]})\circ A_{[k]}+\left(I-{\rm diag}([S_N(X_{[k]})\circ A_{[k]}]{\mathbb 1})\right). 
\end{equation}
This expression can be understood as zeroing out the entries in the similarity matrix $S_N(X_{[k]})$ by element-wise multiplication with $A_{[k]}$, and then adding the appropriate value to the diagonal (the new vector of row-sums is now given by $[S_N(X_{[k]})\circ A_{[k]}]{\mathbb 1})$, so the result becomes row-stochastic.  

With $W$ defined, the meaning of the dynamics in (\ref{eq:newX}) can be better understood.  The $i^{th}$ agent's new opinion on \mbox{topic $j$} is the convex combination of agent $i$'s neighbors' opinions on the topic, and its own, where more weight is given to neighbors with opinions that align better with those of \mbox{agent $i$} {\it over all topics}.  That is to say, in this model, agents are more receptive to the opinions of those who's overall opinion profile mirrors their own, reinforcing each other's point of view.  Also, agents with lots of connections will tend to be more influenced by their (many) friends, while those with fewer connections will put more stock in their own ideas.

\begin{figure}[t]
      \centering
      {\text{Natural Formation of Echo Chambers}}\\[1ex] 
    \includegraphics[width=3.2in]{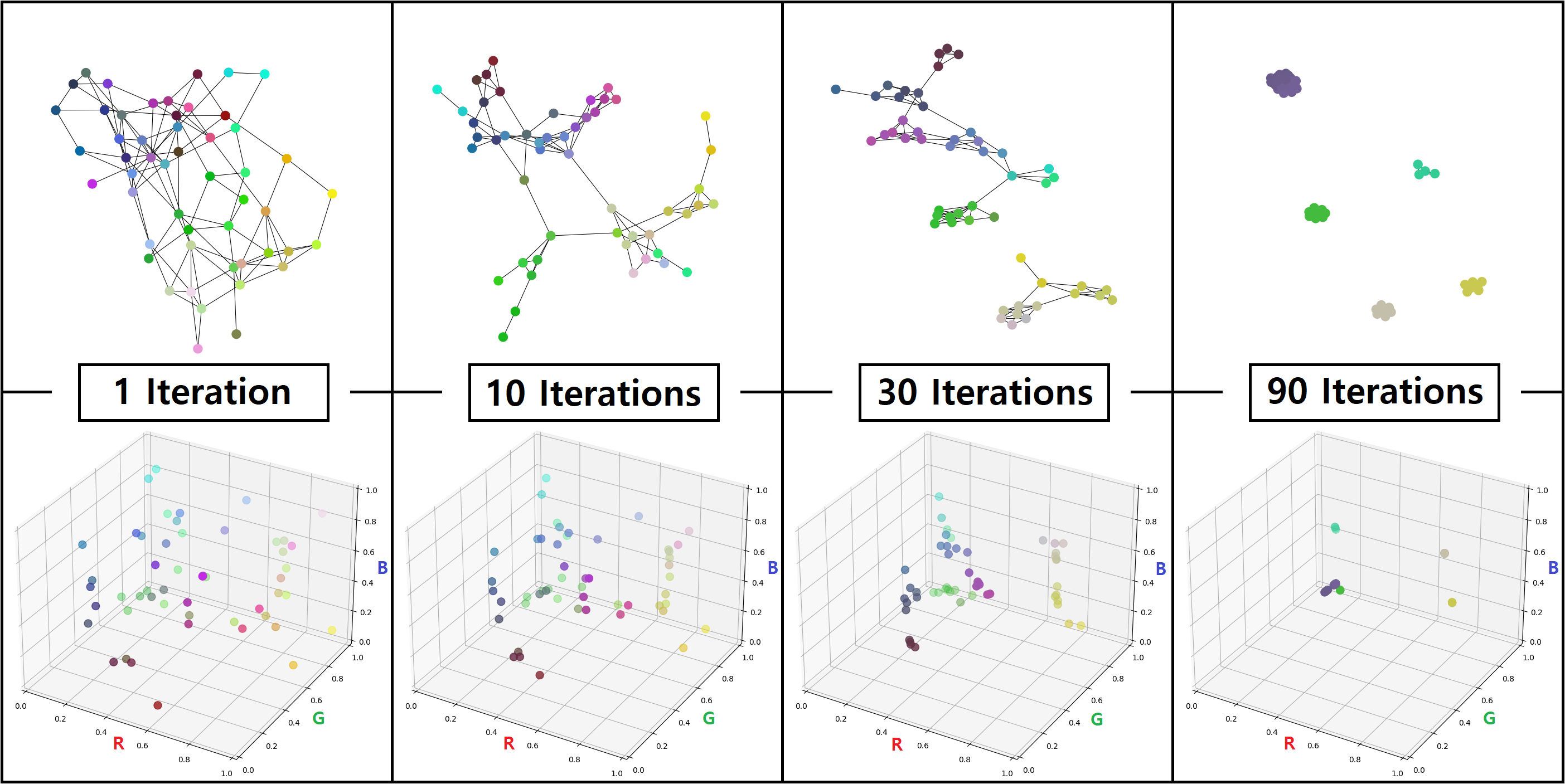}
      \caption{The nonlinear, stochastic Social Network Model in (\ref{eq:newX}) and (\ref{eq:nextA}) results in strong homophily, where agents congregate with those of similar opinions and reject differing opinions.}
      \label{fig:echoChambers}
\end{figure}

Besides the dynamics in (\ref{eq:newX}), which describe the evolution of a vector of opinions associated with each {\it vertex}, or agent, in the social network, our network model also considers the co-evolution of {\it edges}, or connections, among agents in the social network graph.  Like other {\it bounded confidence} models \cite{proskurnikov2017tutorial,proskurnikov2018tutorial}, we describe a situation where agents are insensitive to the opinions of other agents with sufficiently different opinion profiles.  While other models explicitly define a threshold characterizing the degree to which opinions between agents can differ while remaining connected, our notion is stochastic, where connections are randomly sampled from an evolving probability distribution.

In particular, we consider the matrix:
\begin{equation}\label{eq:similar_power}
    \hat{S}_{[k]}:={\cal R}(S_N^{\circ \theta}(X_{[k]}))
\end{equation}
where \mbox{$\theta\in\mathbb Z^+$} is a positive integer modeling parameter, and the notation $M^{\circ \theta}$ describes the operation of raising each element of a matrix $M$ to the $\theta^{th}$ power (or Hadamard multiplying the matix M $\theta$ times, $M\circ M\circ\dots\circ M$). Raising the entries in a stochastic matrix to a positive power $\theta$ and then renormalizing has the effect of driving the larger entries closer to $1$ and zeroing out the smaller entries, making the distinctions between values more extreme.  $A$ then evolves as:
\begin{equation}\label{eq:nextA}
    a_{ij[k+1]}=a_{ji[k+1]}=
    \begin{cases}
    1, & \text{if } \gamma\sim U_{\{0,1\}} < \text{max}(\hat s_{ij[k]},\epsilon) \\ & \text{and } i\ne j
\\
    0, & \text{otherwise}
    \end{cases}
\end{equation}
where $\gamma$ is a sample from $U_{\{0,1\}}$, the uniform distribution on the unit interval, and $\epsilon<<1$ is a modeling parameter establishing the minimum probability for edge formation.  Note that because $\hat{S}_{[k]}$ is a stochastic matrix, its entries are non-negative and bounded by $1$, so edge formation occurs by randomly sampling the uniform distribution and checking to see if the sample is less than the associated entry in $\hat{S}_{[k]}$---or $\epsilon$ (in case the associated entry in $\hat{S}_{[k]}$ has become very small or zero, $\epsilon$ ensures that there is always {\it some} chance of edge formation, as occasionally agents may form connections in spite of strong differences of opinion). The dynamics in (\ref{eq:newX}) and (\ref{eq:nextA}) thus describe our nonlinear, stochastic model of opinion formation and connection co-evolution, respectively.

\section{Three Controller Archetypes}\label{sec:controllers}
The model illustrated in Section \ref{sec:model} provides a network composed of one specific type of a user on social media (i.e. the standard agent), one who desires to connect with those agents who are most similar. To better understand how influential bodies drive opinions in social networks, we present three controller archetypes modeled off inluencer behaviors seen in real life social networks that attempt to drive opinions of the network agents in different ways: 

\begin{itemize}
    \item {Stubborn Agent:} Attempts to move the distribution of opinions towards its own opinion value by refusing to alter its opinion.
    \item {Popular Agent:} Attempts to shape the distribution of opinions, choosing what opinions should become popular.
    \item {Strategic Agent:} Attempts to move the distribution of opinions towards a goal opinion by persuading those furthest from the goal towards the goal.
\end{itemize}

The stubborn agent is included as a standard controller for comparison due to its use in many opinion spread studies. Our strategic and popular agents make use of Hadamard powers in determining where to set their opinions each time-step (see Sections \ref{sec:popularA} and \ref{sec:strategicA}). The selection of the Hadamard power value, $\rho$, creates a spectrum of  behaviors for those two archetypes. Figure \ref{fig:3archetypes} shows a tree representing the relationships between the three controller archetypes. The following subsections provide detailed explanations of each.

Within the subsections for each controller, we do not provide proofs of stability or convergence. As A. Proskurnikov and R. Tempo explain in section 5.3 of \cite{proskurnikov2018tutorial} when talking about HK and DW models (of which ours can be seen as a relative):
\begin{quote}
    ``In spite of many numerical
results and experimental observations, dealing with
the behavior of the... model and its modifications
over complex networks, the compound
of randomness and nonlinear dynamics makes these
models very hard for mathematical investigation."
\end{quote}
Proofs of this nature for networks and controllers like what we present in this paper are often long and complex. Due to the limited space, we leave our proofs of convergence and stability for subsequent work.

\subsection{Stubborn Agent}

The goal of the stubborn agent is to move the opinions of the network towards its own opinion. It does this by retaining its opinion each time-step. In this way, whenever an edge is created between it and another agent, the stubborn agent's opinion becomes part of the convex combination that determines that agent's opinion in the next time-step. The stubborn agent, however, is not influenced by that other agent. It acts as in immutable input signal into the network.

The update steps for the stubborn agent differ slightly from the standard agent. In Equation \ref{eq:weights}, we set its row in $W(X_{[k]},A_{[k]})$ to zeros and its diagonal element to one. This results in its row in the weight matrix having full weighting to itself and none to any other agent. Or in other words, when we calculate $W_{[k+1]}$ in Equation \ref{eq:newX}, 
$x_{stubborn[k+1]}^\top = x_{stubborn[k]}^\top$.

\subsection{Popular Agent}\label{sec:popularA}
The popular agent does not have an opinion of its own but instead chooses which opinions among its neighbors it should propagate. To do this, it bases its opinion completely on its neighbors, weighted according to the similarity of each neighbor's opinion to every other neighbor's opinion. 

\begin{definition}
Let $i$ be the row index of the popular agent in the adjacency matrix, $A$. The set of neighbor indices, $\mathcal{N}^i$, is given by:
    \[
    \mathcal{N}^i:=\{j \in V \;|\; a_{ij}=1\}, a_{ij} \in A
    \]
\end{definition}

\begin{definition} \label{def:neighbor}
The $j^{th}$ element of the {\em neighbor distance vector}, $d^i_{[k]}$, for agent, $i$, at time, $k$, and its corresponding {\em neighbor weight vector}, $\mathbf{{\omega}}_{[k]}$, are given by:
    \[
    \begin{array}{rcl}
    d^i_{j[k]} := \sum_{l\in \mathcal{N}^i,l\ne j} ||x^{\top}_l - x^{\top}_j||,&{\rm and}&
    \mathbf{{\omega}}_{[k]} := {\cal R} (d^{i\top}_{[k]}).
    \end{array}
    \]
\end{definition}

The resulting stochastic weight vector, $\mathbf{{\omega}}_{[k]}$, contains weights for each neighbor of the popular agent based on the difference of opinion of each neighbor from every other neighbor. The more similar to all other neighbor opinions, the lower the weight value. The more different, the larger the weight value.

\begin{definition}
    The {\em emphasized neighbor weight vector} increases the popular agent's emphasis on the level of similarity of the opinions by taking the $\rho$th Hadamard power of $\omega_{[k]}$ and re-normalizing: 
\begin{equation}\label{eq:rho_weight}
    {\hat \omega}_{[k]} :={\cal R}({\omega}^{\circ \rho}_{[k]})
\end{equation}
\end{definition}

\begin{definition}\label{def:subX}
    $X^i_{[k]}$ is the sub-matrix of $X_{[k]}$ that only contains rows corresponding to popular agent, $i$. 
\end{definition}

With these definitions, the opinion update of the $i^{th}$ row of $X_{[k]}$ represented by the popular agent is given by:

\begin{equation}\label{eq:popularX}
    x_{i[k+1]} = \hat{\omega}_{[k]}X^{i}_{[k]}
\end{equation}

The resulting opinion of the popular agent will not necessarily be the average of the neighboring opinions, but instead will be shifted either towards those opinions that are most popular among its neighbors or those that are fringe opinions among its neighbors, depending on the value of $\rho$ in the Hadamard power of Equation \ref{eq:rho_weight}. 

\begin{fact}\label{theorem:pop1}
As $\rho \rightarrow -\infty$, $x_{[k+1]}$, the opinion of the popular agent at time-step $k+1$, will become the average of those agents with maximum opinion similarity to all other neighboring agents at time $k$.
\end{fact}

\begin{fact}\label{theorem:pop2}
As $\rho \rightarrow 0$, $x_{[k+1]}$, the opinion of the popular agent at time-step $k+1$, will become the average opinion of all neighboring agents at time $k$.
\end{fact}

\begin{fact}\label{theorem:pop3}
As $\rho \rightarrow \infty$, $x_{[k+1]}$, the opinion of the popular agent at time-step $k+1$, will become the average of those agents with minimum opinion similarity to all other neighboring agents at time $k$.
\end{fact}

The proofs of Facts \ref{theorem:pop1}--\ref{theorem:pop3} are trivial and stem directly from the behaviors of renormalized Hadamard powers of stochastic matrices explained just before Equation \ref{eq:nextA}. These three facts mean that by selecting values for $\rho$, we can determine how the popular agent views the collective opinions of its neighbors. By having a large, positive $\rho$, the popular agent attempts to drive the collective opinions of its neighbors towards the fringe opinions, or in other words it tries to make those fringe opinions more popular, giving it the title, ``Popularizer." By making $\rho$ large and negative, the popular agent attempts to drive opinions to strengthen the collectively most popular opinion, giving it the title, ``People Pleaser." At zero, the popular agent tries to drive opinions towards the average of its neighbors, giving it the title, ``Conciliator."

\subsection{Strategic Agent}\label{sec:strategicA}
The purpose of the strategic agent is to direct the network towards some goal opinion by basing its opinion off its neighbors and coaxing them towards the goal. 
\begin{definition}
Let $i$ be the row index of the strategic agent and $g^i$ be the goal opinion of the strategic agent. Using Definition \ref{def:neighbor} Section \ref{sec:popularA} to define a neighbor set, the {\em neighbor distance vector} for the strategic agent at time, $k$, is given by:
\begin{equation}\label{eq:stratD}
    d_{j[k]} := ||x^{i\top}_{j[k]} - g^i|| \text{ for } j\in \mathcal{N}^i
\end{equation}
where $x^{\top}_{j[k]}$ denotes the $j^{th}$ row of $X_{[k]}$.
\end{definition}

Since we do not know what the resulting distances will be at any time-step, we cannot arbitrarily choose a set weighting for the goal opinion for the convex combination. Instead, we append to the bottom of $d$ the minimum value of $d$, meaning that the weighting of the goal opinion will be the same as the closest of the neighbors to it after normalization. We can now use second half of Definition \ref{def:neighbor} to normalize by the vector sum to get the weight vector, $\mathbf{\omega}_{[k]}$.

Using Equation \ref{eq:rho_weight} from Section \ref{sec:popularA}, we can increase $\rho$ to further weight towards the most distant opinion and away from the goal opinion whose weight matches the minimal element of $\mathbf{\omega}_{[k]}$. We once again defin $X^i_{[k]}$ as the sub-matrix of $X_{[k]}$ according to Definition \ref{def:subX}. We transpose and append the goal opinion, $g^i$, at the bottom row of $X^i_{[k]}$ so that the matrix dimensions for both $X^i_{[k]}$ and $\mathbf{\hat\omega}_{[k]}$ are compatible in Equation \ref{eq:popularX}. We can now use that same opinion matrix update function 
to update the opinions of the strategic agents in $X_{[k+1]}$. So long as $\rho \ne \infty$, each convex combination will include influence from $g^i$, nudging standard agents towards the goal opinion.

\begin{figure}[t]
  \centering
  \text{Initial Network}\\[1ex] 
  \begin{subfigure}[t]{1.625in}
    \centering
    \includegraphics[width=\textwidth]{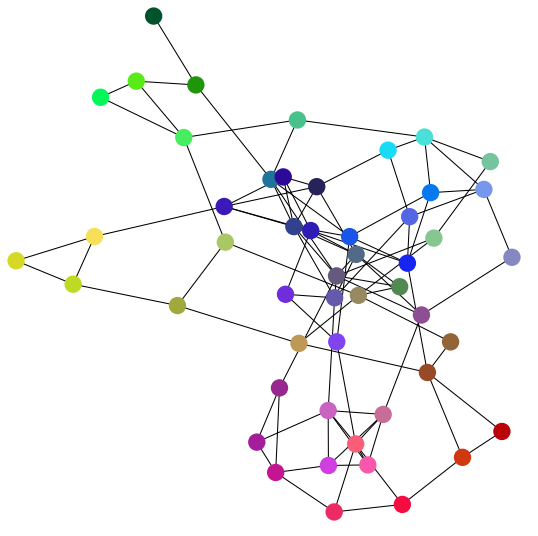}
    \caption{Colored Opinions}
    \label{fig:initial_opinions_color}
  \end{subfigure}
  \hfill 
  \begin{subfigure}[t]{1.625in}
    \centering
    \includegraphics[width=\textwidth]{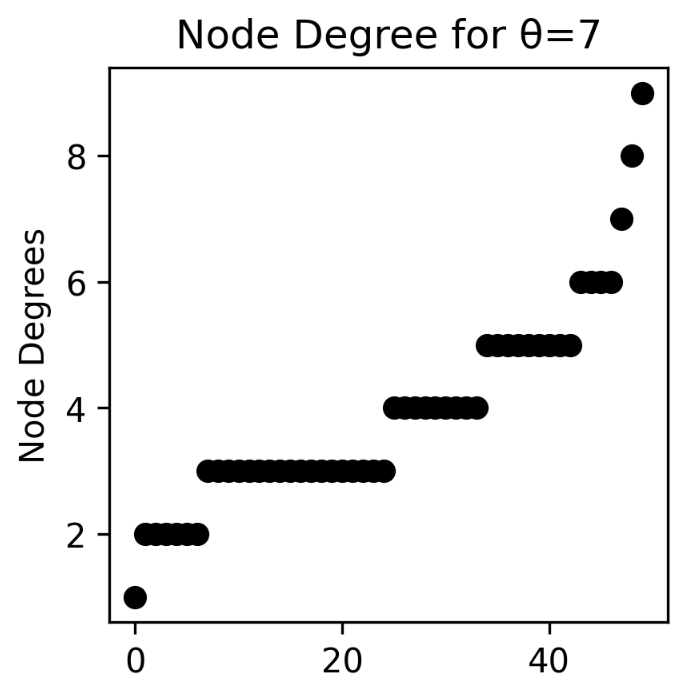}
    \caption{Node Degree}
    \label{fig:initial_node_degree}
  \end{subfigure}
  \caption{Initial random network of 50 standard agents used for controller archetype experiments. Each agent has 3 opinions, and $\theta=7$ in Equation \ref{eq:similar_power} for edge connectivity.
      Average opinions once stable are $[0.48, 0.44, 0.52]$.}
  \label{fig:initial_network}
\end{figure}

\begin{figure*}[t]
  \centering
  \text{Results of Controller Archetype Simulations}\\[1ex] 
  \begin{subfigure}[t]{0.32\textwidth}
    \centering
    \includegraphics[width=\textwidth]{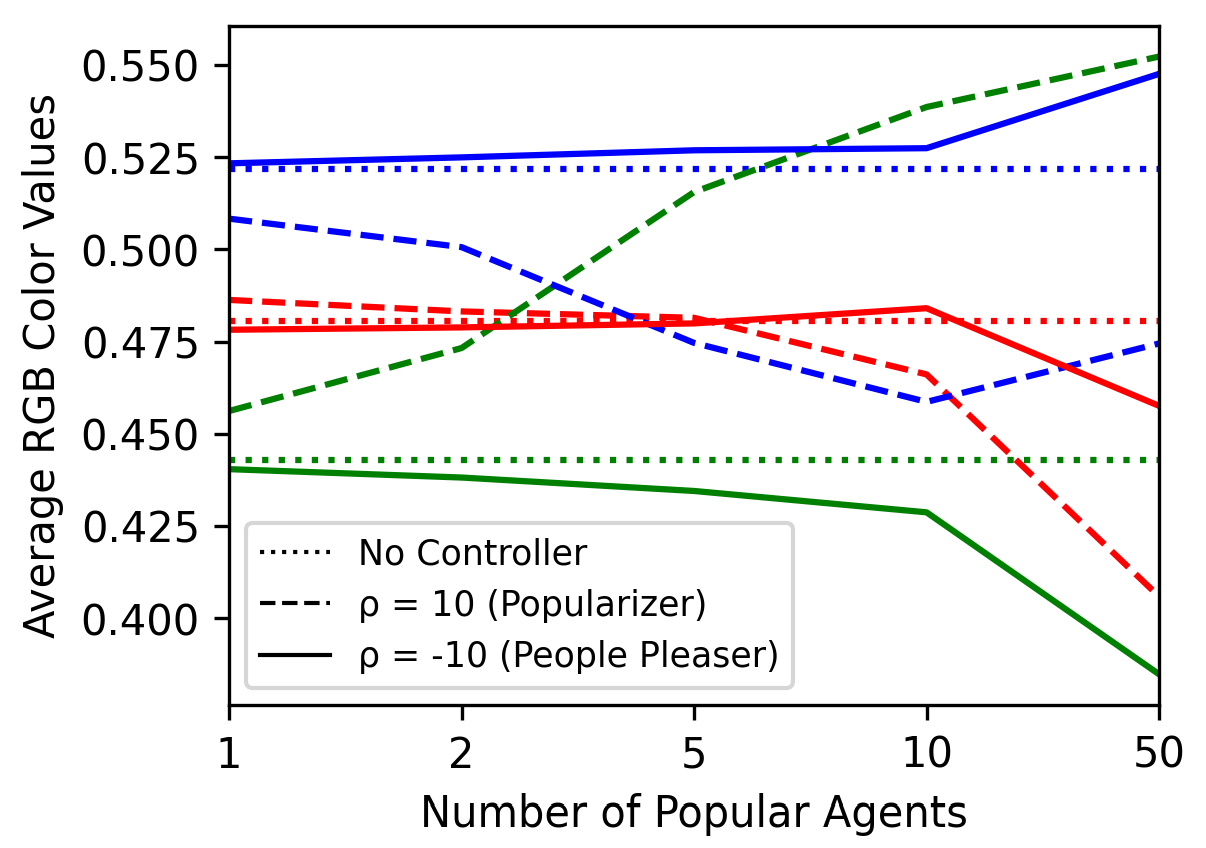}
    \caption{Resulting average RGB opinion values for different numbers of ``people pleasers" and ``popularizers."}
    \label{fig:popular_spectrum}
  \end{subfigure}
  \hfill
  \begin{subfigure}[t]{0.31\textwidth}
    \centering
    \includegraphics[width=\textwidth]{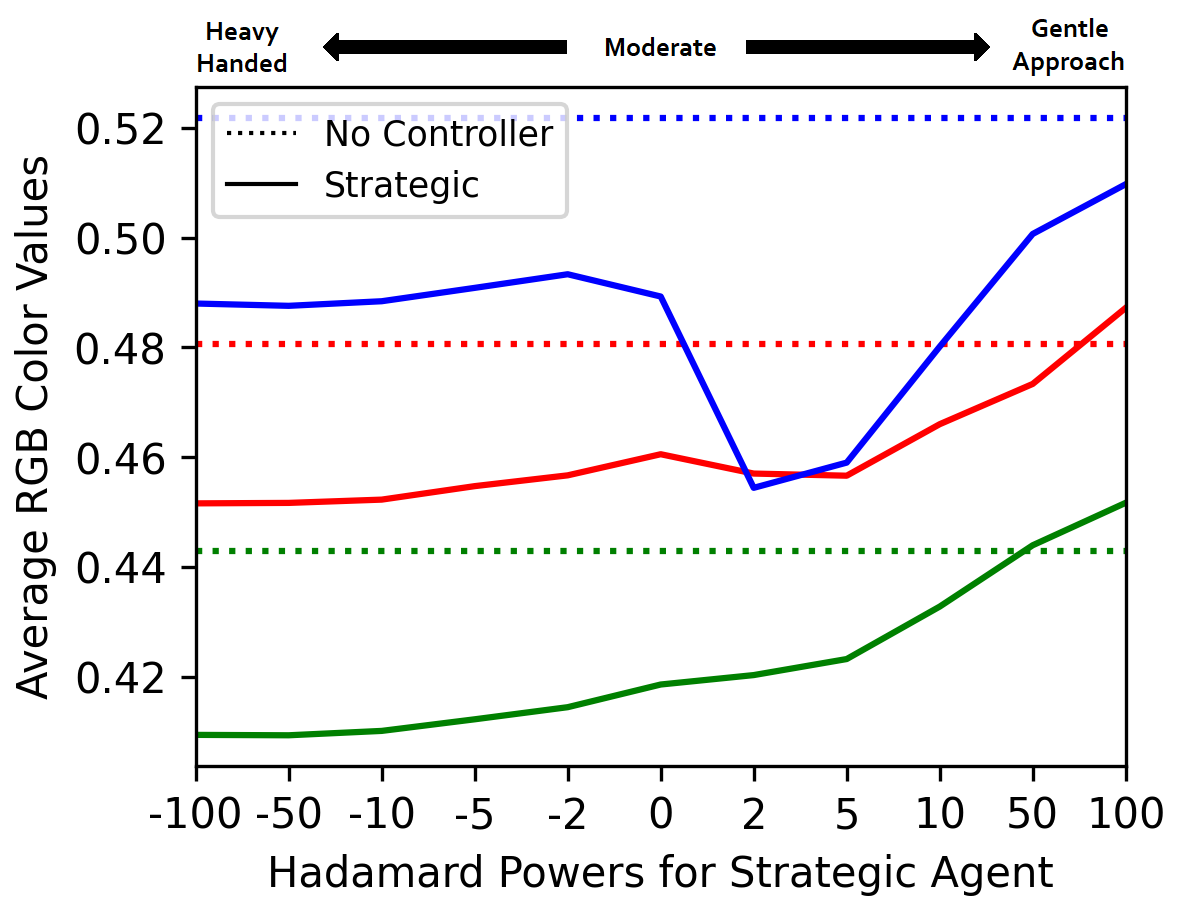}
    \caption{Effects of different Hadamard powers for the strategic agent with goal opinion of $[0,0,0]$.}
    \label{fig:Strategic_spectrum}
  \end{subfigure}
  \hfill
  \begin{subfigure}[t]{0.28\textwidth}
    \centering
    \includegraphics[width=\textwidth]{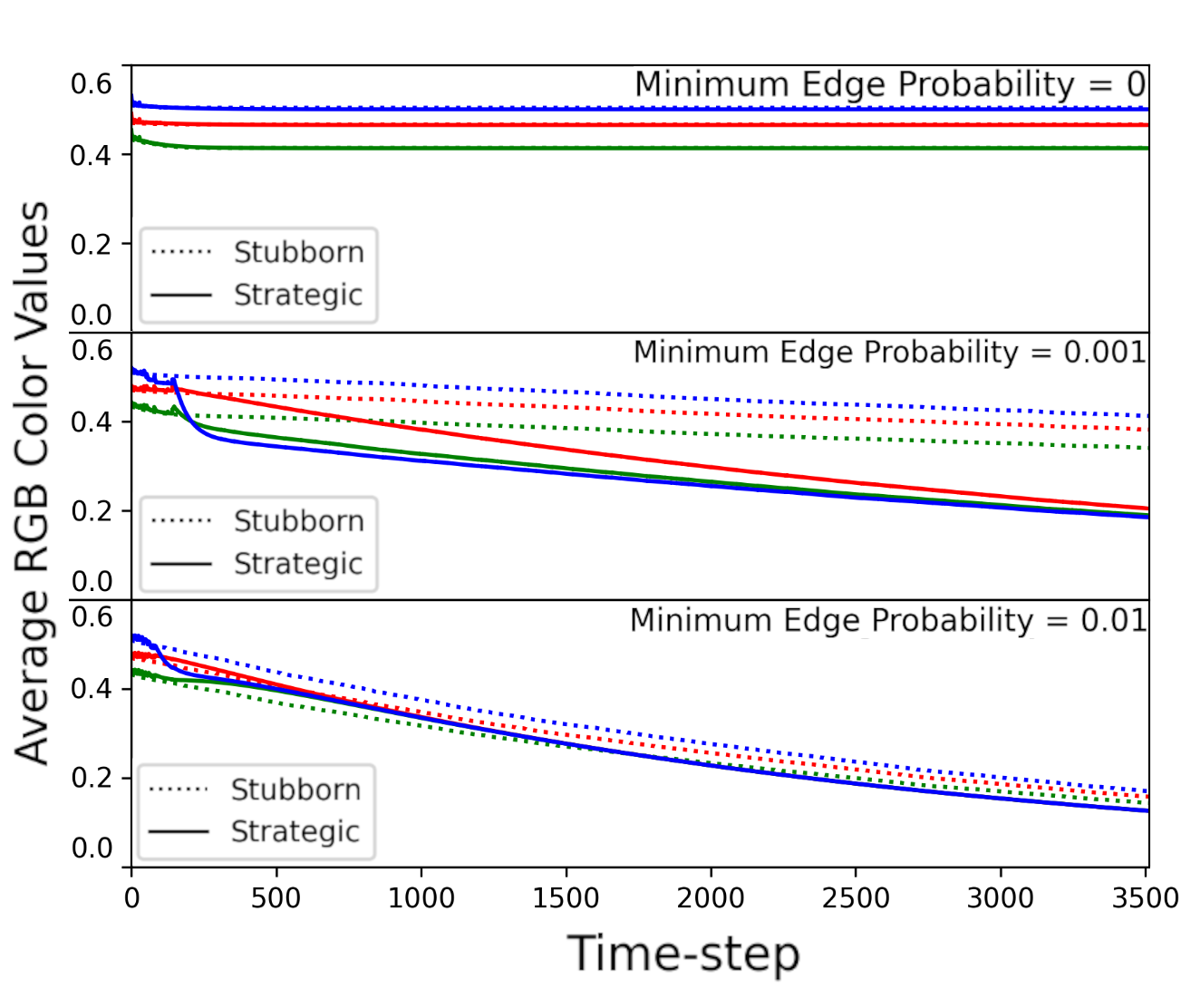}
    \caption{The comparison between strategic and stubborn agents with target opinion of $[0,0,0]$ given different minimum edge probabilities.}
    \label{fig:strat_v_stub}
  \end{subfigure}
  \caption{Overview of simulations demonstrating the influence of various agents in opinion dynamics. Each figure represents a unique setup and outcome, illustrating the complex interplay between different agent strategies and their effects on opinion distribution within a network. The results of Figure \ref{fig:popular_spectrum} are explained in Section \ref{subsec:popular}, Section \ref{subsec:strat_spectrum} for Figure \ref{fig:Strategic_spectrum}, and Section \ref{subsec:stratVSstub} for Figure \ref{fig:strat_v_stub}.}
  \label{fig:combined_figures}
\end{figure*}

\section{Experiments and Results}\label{sec:results}

Figure \ref{fig:initial_network} shows the initial 50 agent network used for all proceeding experiments. We set $m=3$ (three opinions) so that we can visualize opinions by assigning each opinion to a respective RGB value. For edge connectivity, we set $\theta=7$ in Equation \ref{eq:similar_power} so that as time progresses, agents isolate from each other into groups of similar opinions (echo chambers). The average RGB opinions of this network once stability is reached are  $[0.48, 0.44, 0.52]$. The most common opinions of this network are a high-blue and low-green. The following experiments and tests are to see how each controller archetype changes the resulting final average opinion values of the network.

\subsection{Popular Agent Spectrum}\label{subsec:popular}
We envision popular agents as any number of influencer accounts that standard agents may choose to follow. The popular agents then
base their opinions on their followers and act as independent input signals into the system (no edges to each other).
We conduct three experiment sets. First, we run until network stabilization without popular agent influence as a control comparison. Next, we set the Hadamard power, $\rho=-10$, in Equation \ref{eq:rho_weight} of Section \ref{sec:popularA} to propagate the dominant opinions of the network (``people pleaser"). Lastly, we set $\rho=10$ to increase the influence of the fringe opinions of the network (``popularizer"). For the second and third experiment sets, we run simulations for 180 time-steps using 1, 2, 5, 10, and 50 popular agents with the same $\rho$ value.

Figure \ref{fig:popular_spectrum} shows the resulting average RGB opinion values. As expected, the ``people pleasers" increase the blue opinion while decreasing the green opinion and eventually the red, although the influence of the people pleasers is minimal when there are few of them. The ``popularizers," however, show a stronger influence on the opinions of the network even when only one popular agent is present. With only five popularizers, the formerly unpopular opinion of a high-green becomes dominant over a high-blue opinion. However, once enough popularizers are present, the high-blue opinion eventually becomes fringe enough that some of them take that stance and propagate it through the network, which is why we see an increase in blue opinion at fifty popularizers.

\subsection{Strategic Agent Spectrum}\label{subsec:strat_spectrum}

To explore the impact of the strategic agent spectrum on network opinions, we set a strategic agent's goal to $[0,0,0]$ and conducted eleven experiments with Hadamard power values from $-100$ (heavy-handed) to $100$ (gentle approach), as depicted in Figure \ref{fig:Strategic_spectrum}. A heavy-handed strategy aligns the strategic agent's opinion closely with its goal, impacting connectivity with distant standard agents by increasing the likelihood of losing connections. Conversely, a gentle approach aligns the strategic agent's opinion more closely with distant agents, diminishing the goal's influence. These experiments, after 180 iterations, reveal that a heavy-handed approach lowers RGB values, moving them closer to zero, but at $\rho=2$ and $\rho=5$, the blue opinion significantly drops. The gentle approach balance at $\rho=2$ and $5$ enables the strategic agent to maintain connections effectively while exerting enough influence towards the goal opinion.

\subsection{Strategic Agent Versus Stubborn Agent} \label{subsec:stratVSstub}

Both strategic and stubborn agents move the network's opinion distribution toward a target opinion---stubborn agents aim for their own opinion, while strategic agents target a predefined goal without directly adopting it. Given the network's dynamic nature, an edge from a standard agent to a strategic or stubborn agent is not always present. The strategic agent's tactic of adjusting its opinion based on acquired neighbors is crucial for influencing the network's opinions through enhancing the likelihood of maintaining connections once established.
To compare these two agents, we test various edge creation probabilities, $\epsilon$, in Equation \ref{eq:nextA}, over 3500 time-steps with either a stubborn or strategic agent in the network, both targeting $[0,0,0]$. The strategic agent's $\rho$ is set to 2, following outcomes from Section \ref{subsec:strat_spectrum}.

Figure \ref{fig:strat_v_stub} shows the resulting average RGB opinion values for both stubborn and strategic agents with minimum edge probabilities of zero, 0.001, and 0.01. 
At zero, the opinion of $[0,0,0]$ eventually becomes the furthest from any regular agent, and after applying the Hadamard power and normalizing before Equation \ref{eq:weights}, the weight of influence becomes effectively zero.
When a minimum edge probability of 0.001 is used, the strategic agent can gain followers, adopt opinions similar to theirs, and retain edges, outperforming the stubborn agent. The stubborn agent's opinion remains starkly different and must rely solely on the minimum edge probability to facilitate influence.
At an edge probability of 0.01, the stubborn agent is comparable to the strategic agent, though still at a slight disadvantage. Overall, the strategic agent has an advantage in moving the opinions of agents in the network towards its target when the probability is low but not zero due to its ability to adopt opinions similar to the standard agents in the network.

\begin{figure}[t]
      \centering
      {\text{Experimental Results for Opinion Inference }}\\[1ex] 
    \includegraphics[width=3.2in]{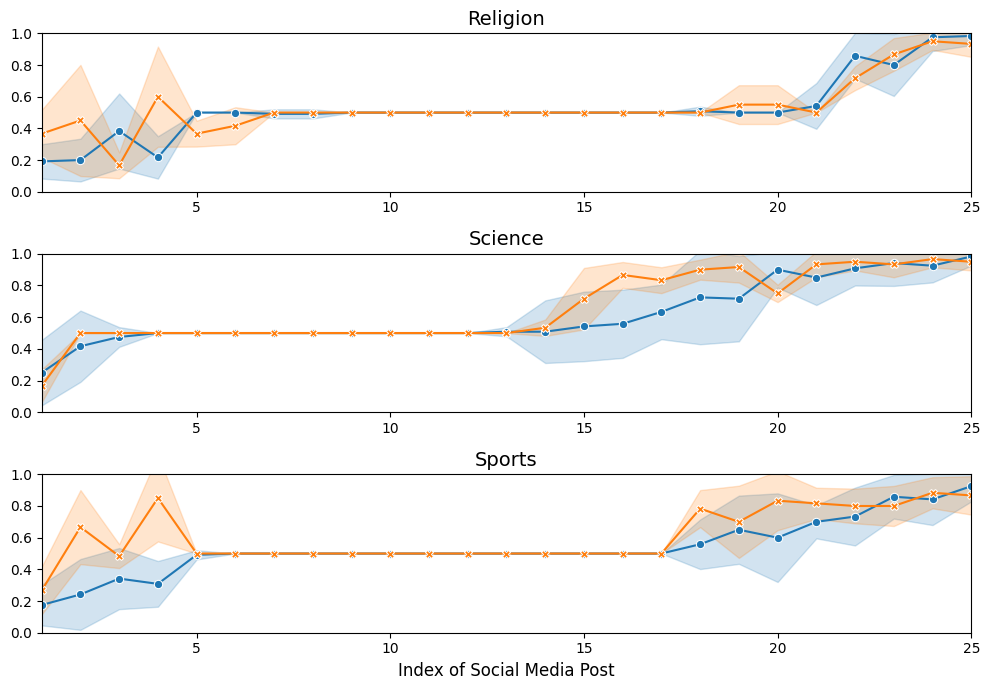}
      \caption{Rankings by human annotators (Orange) and GPT-4 (Blue) of 25 social media posts about Religion, Science, and Sports on a scale from zero (oppose) to one (support). Variance is low when ranking opinions with strong language supporting a topic, but both humans and GPT-4 have some difficulty with negative language due to some posts with sarcasm, which were included in the study. Posts with neutral opinions or without reference to a topic were nearly universally identifiable by both humans and GPT-4.}
      \label{fig:post_rankings}
\end{figure}

\section{Generative AI Makes Implementation of Opinion Controllers Easy}\label{sec:LLM}
In this section, we investigate how an automated agent can drive opinions in social networks, not as a how-to demonstration but to emphasize that now is the time to think of the ethics and real-life implications of opinion and social control research in control theory. We begin by first showing than an LLM can act as an opinion inference engine for social media posts, and then we show that generative AI can create content based on a provided opinion vector. The feedback control relationship can be seen in Figure \ref{fig:ControlDiagram}.

\subsection{Opinion Inference Engine}
\begin{figure}[t]
  \centering
  \text{Memes on Taking Action Against Climate Change}\\[1ex] 
  \begin{subfigure}[t]{1.625in}
    \centering
    \includegraphics[width=\textwidth]{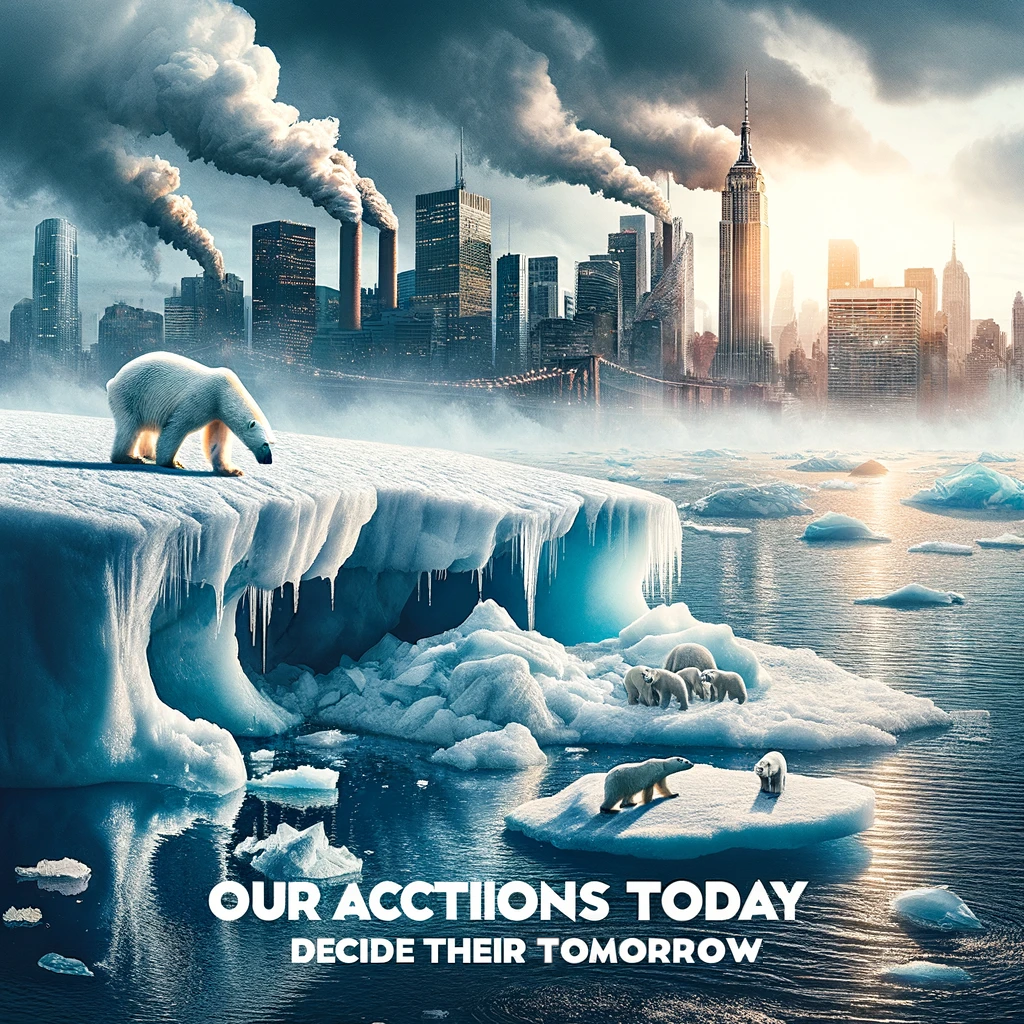}
    \caption{Opinion Value 1.0 (Support)}
    \label{fig:climate1a}
  \end{subfigure}
  \hfill 
  \begin{subfigure}[t]{1.625in}
    \centering
    \includegraphics[width=\textwidth]{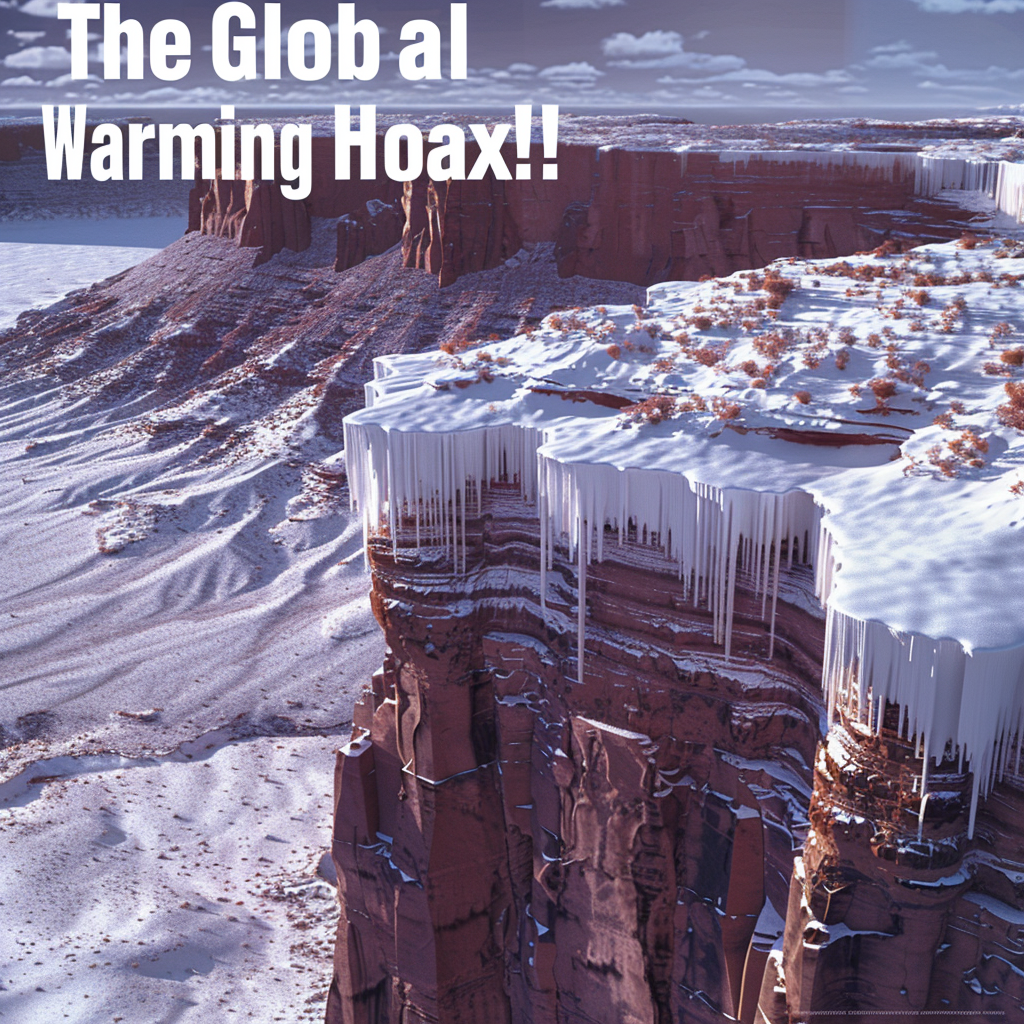}
    \caption{Opinion Value 0.1 (Oppose)}
    \label{fig:climate1b}
  \end{subfigure}
  \caption{Memes generated by Dalle-3 (left) and Midjourney V6 (right). Notice slight errors in the text spelling and spacing, but, nevertheless, these demonstrate that there are multiple tools available to automate effective meme generation responding to a numerical scale on an issue.}
  \label{fig:climate_both}
\end{figure}

For the first experiment, we gathered from Facebook, Twitter, and Reddit 25 social media posts on three topics of religion, science, and sports, and included posts that were not on any of those topics as a control. We had twelve human volunteers annotate the opinion vectors for each individual opinion topic. We then had six independent instances of GPT-4 \cite{OpenAI_GPT4_2023} output its inference on the opinion vectors in $\mathbb{R}^3$. Figure \ref{fig:post_rankings} shows the variance for the human annotators (orange) and GPT-4 (blue). We see a general correlation between opinion assessments, but where the variance is low in posts with strong supportive language on a topic, there is some disagreement in strong negative language due to sarcasm (some sarcastic posts were included in the data). Neutral stances on topics and posts that were not on any of the three topics were nearly universally identifiable by both humans and GPT-4. After averaging the opinion rankings for the human and GPT-4 assessments, we find that the difference in opinions on average was 0.023 for religion, 0.177 for science, and 0.11 for sports. These results show that the ability of GPT-4 to assess opinion vectors of social media posts is fairly similar to human assessments, though not equal. The capability of GPT-4 to take in image data means these results may extend to interpretation of image content as well. For example, GPT-4 properly identified the opinion values of the images in Figure \ref{fig:climate_both}. 

\begin{table*}[ht]
\caption{Chat-GPT Generated Media Post Given Controller-Generated Opinion Value}
\label{tab:chatGPT}
\centering
\begin{tabular}{||M{0.1\linewidth}|M{0.8\linewidth}||}
\hline
Opinion Value & Social Media Bot Posts About Taking Action on Climate Change\\
\hline\hline
1.0 (Support) & 

I simply cannot fathom why anyone would oppose the fact that climate change is real. Ignoring this issue is not an option. We need to come together as a global community to address this challenge head-on by reducing our carbon footprint, investing in renewable energy, and advocating for policies that protect our planet for future generations.\\ 
\hline
0.7 & 
We have a responsibility to take care of our planet and ensure that it remains habitable for future generations.
Despite the challenges we face, I'm optimistic about the steps we can take to address climate change. Renewable energy sources like wind and solar power are becoming more affordable and accessible every day, and innovative technologies are helping us to reduce our carbon footprint and conserve resources.\\
\hline
0.5 (Neutral) & Climate change is a complex and often controversial issue, with arguments for and against its existence and significance. 
On the one hand, there is a wealth of scientific evidence that suggests that
human activities are a major contributing factor. 
There are those who dispute these findings and
contend that efforts to address climate change may be too costly or difficult to implement, and that more research is needed before we can take action.\\
\hline
0.3 & 
I also think that some of the proposed solutions to address climate change may be unrealistic or unfeasible, particularly for those who live in developing countries or who rely on certain industries for their livelihoods. While I believe in taking care of our planet, I think we need to balance this with practical considerations and not make hasty decisions that could have unintended consequences.\\
\hline
0.1 (Oppose)& I simply cannot stand the constant fearmongering and doom-and-gloom surrounding the topic of climate change. 
I think that the push for climate change solutions often overlooks the economic and social costs of these policies. It's not as simple as just ``going green" --- there are real trade-offs and sacrifices that would need to be made, and I don't think it's fair to expect everyday people to bear the brunt of these changes. \\
\hline
\end{tabular}
\end{table*}

\subsection{AI Generation from Opinion Vectors}
To show that generative AI can create posts based on opinion vectors, we provided GPT-4 with the topic of ``taking action on climate change" and had it create short posts following opinion values from 0.1 to 1.0. Table \ref{tab:chatGPT} shows the posts generated by Chat-GPT, revealing the subtle nuances in stances toward climate change action even between similar opinion values. 
Although current image generation AI do not have the full capability of simply taking in an opinion vector and topic, interpret it, and then generate a resulting image following that opinion, we can use LLMs like GPT-4 to generate an image prompt for image AI like Dalle-3 and Mdjourney V6. Figure \ref{fig:climate_both} shows two examples of memes generate by Dalle-3 (left) for support and Midjourney V6 (right) for opposition against climate change action. With the vast improvements in video generation as seen in OpenAI's Sora \cite{liu2024sora}, it will not be long before video content can be generated in the same manner.

\section{ Ethical Considerations 
}\label{sec:ethics}
Although previous work on opinion control may have considered the idea in the abstract, this paper demonstrates how easy it now is to implement such controllers on real social media networks using current generative AI technologies.  This widespread access to systematic methods for mass manipulation highlights the urgent need for discussions on the morality of applying control methods to people and the development of corresponding ethical principles.

Certainly there are many uses for automated agents in social media networks that would seem to contribute to the common good, such as compensating for levels of homophily that may appear to be unhealthy, or making introductions between agents with common interests.  Nevertheless, the same control methods that enable these capabilities can just as easily contribute to real social harm \cite{debuse2023automatic, mosleh2022measuring, bovet2019influence}. 

This isn't the first time the scientific community has  conducted research or developed technologies with the potential for real social harm, however, so there are frameworks we can use for considering the ethical implications.  For example, the Belmont Report \cite{united1978belmont}
offered critical guidance for biomedical research in 1978, and the Menlo Report \cite{bailey2012menlo} offered similar principles for research on information and communications technologies in 2012:
\begin{enumerate}
    \item Respect for Persons, including Informed Consent,
    \item Beneficience,
    \item Justice, and
    \item Respect for Law and Public Interest, including Transparency    
\end{enumerate}

Kevin Macnish and Jeroen van der Ham have more recently modified the Menlo Report \cite{macnish2020ethics}, with a special focus on cybersecurity research that includes a section looking beyond research activities to explore the ethics of cybersecurity development in industrial contexts.  This application closely parallels questions surrounding the ethics of publishing effective opinion or social control techniques since such techniques are both research and, as demonstrated here, nearly immediately deployable in practice using widely available generative AI.

The Association for Computing Machinery (ACM), however, developed a framework in 2022 even more applicable to social control than cybersecurity in their  ``Statement on Principles for Responsible Algorithmic Systems" \cite{ACM2022statement}:
\begin{enumerate}
    \item Legitimacy and Competency,
    \item Minimizing Harm,
    \item Security and Privacy,
    \item Transparency,
    \item Interpretability and Explainability,
    \item Maintainability,
    \item Contestability and Auditability,
    \item Accountability and Responsibility,
    \item Limiting Environmental Impacts
\end{enumerate}
These principles correspond strongly to those from \cite{bailey2012menlo} and refined in \cite{macnish2020ethics}, but add new refinements such as interpretability/explainability and contestability/auditability. 

Since then, various organizations have worked on further refinements in the context of ethical AI systems, resulting in guidelines for Trustworthy and Responsible AI from the National Institute of Standards and Technology (NIST) \cite{NISTethics}, that added fairness to its list of principles for ethical systems, and various other guidelines, of which the list from Intel is typical \cite{IntelEthics}.  These criteria, however, continue to evolve and could benefit from input from the controls community.    

\section{Conclusion}

Automatic methods designed to change people's minds have reached a new level of expertise.  This paper presented a new model of Social Media Networks along with algorithms for three opinion controllers.  Results from a simulation study were then described, and details for using generative AI to deploy such controllers were illustrated--including experimental results demonstrating the efficacy of using Large Language Models for quantifying the opinion characteristics of social media posts.  Ethical concerns and frameworks from related fields were then presented with an invitation for more work understanding the morality of using automatic control methods for persuasion.   

Organizations like the NSF and DARPA have long recognized the risks of powerful techniques developed for feedback control, including jobs lost to automation \cite{NSFSolicitation2023} and, specifically, the need for cognitive security to protect populations from influence operations and mass manipulation \cite{waltzman2017weaponization}. Future work could explore how to accomplish these goals.

\bibliographystyle{IEEEtran}
\bibliography{BIB}

\end{document}